\documentclass[twocolumn]{aa}  

\usepackage{graphicx}
\usepackage{txfonts}

\usepackage[colorlinks=true, pdfborder={0 0 0}, citecolor=blue]{hyperref}

\usepackage{lscape}
\usepackage{booktabs}

\newcommand{\nsingle}{17}

\newcommand{\nsong}{25}
\newcommand{\Hale}{Hale}

\begin{document} 

\title{The spin-orbit alignment of visual binaries}

\author{A.\ B.\ Justesen\inst{1}\fnmsep\thanks{\email{justesen@phys.au.dk}} \and S.\ Albrecht\inst{1}}

\institute{Stellar Astrophysics Centre, Department of Physics and Astronomy, Aarhus University, Ny Munkegade 120, DK-8000 Aarhus C, Denmark.}

\date{Received \today; accepted \today}

  \abstract
   {The angle between the stellar spin-axis and the orbital plane of a stellar or planetary companion has important implications for the formation and evolution of such systems. A study by \citet{1994AJ....107..306H} found that binaries with separations $a\lesssim 30\,$au are preferentially aligned while binaries on wider orbits are frequently misaligned.
   }
   {We aim to test the robustness of the \citet{1994AJ....107..306H} results by reanalysing the sample of visual binaries with measured rotation periods using independently derived stellar parameters and a Bayesian formalism.
   }
   {Our analysis is based on a combination of data from \citet{1994AJ....107..306H} and newly obtained spectroscopic data from the Hertzsprung SONG telescope, combined with astrometric data from \textit{Gaia} DR2 and the Washington Double Star Catalog. We combine measurements of stellar radii and rotation periods to obtain stellar rotational velocities $v$. Rotational velocities $v$ are combined with measurements of projected rotational velocities $v\sin i$ to derive posterior probability distributions of stellar inclination angles $i$. We determine line-of-sight projected spin-orbit angles by comparing stellar inclination angles with astrometric orbital inclination angles.
   }
   {We find that the precision of the available data is insufficient to make inferences about the spin-orbit alignment of visual binaries. The data are equally compatible with alignment and misalignment at all orbital separations.
   }
   {We conclude that the previously reported trend that binaries with separations $a \lesssim 30\,$au are preferentially aligned is spurious. The spin-orbit alignment distribution of visual binaries is unconstrained. Based on simulated observations, we predict that it will be difficult to reach the sufficient precision in $v\sin i$, rotation periods, and orbital inclination required to make robust statistical inferences about the spin-orbit alignment of visual binaries.
   }

   \keywords{Binaries: visual -- Stars: solar-type --  Protoplanetary disks -- Planets and satellites: dynamical evolution and stability --  Methods: data analysis}

   \maketitle


\section{Introduction}
The formation of binary stars is poorly understood. Disc fragmentation, turbulent fragmentation, and dynamical encounters may all play important roles in the formation of binary and higher-order systems \citep{bate_statistical_2019}. If binary stars form preferentially via disc fragmentation, it is more likely that the stellar spins and protoplanetary discs of each star will be mutually aligned with each other and the binary orbit. On the other hand, if binary star formation occurs primarily through turbulent fragmentation or dynamical interactions, the stellar spin-axes in a binary system may be significantly misaligned with each other, the orbital plane, and the circumbinary disc \citep{2010MNRAS.401.1505B, 2018MNRAS.475.5618B}. An initially aligned protoplanetary disc may become misaligned with the stellar spin-axis due to secular interactions with an inclined binary companion \citep{2012Natur.491..418B}. Hot Jupiters are often found in orbits misaligned with respect to the stellar spin \citep{2010ApJ...718L.145W, 2012ApJ...757...18A}. It is however unclear if these planets formed in misaligned protoplanetary discs or if misalignment occurred after planet formation due to high-eccentricity migration \citep[see e.g.][]{2018ARA&A..56..175D}. The spin-orbit alignment of binary stars is therefore important not only for constraining theories of star formation, but also for understanding the orbital evolution of planetary systems. If wide binary stars are found to be preferentially spin-orbit aligned, it is likely that protoplanetary discs are also well-aligned \citep{2007prpl.conf..395M}. 

\begin{figure}
   \centering
   \includegraphics[width=\columnwidth]{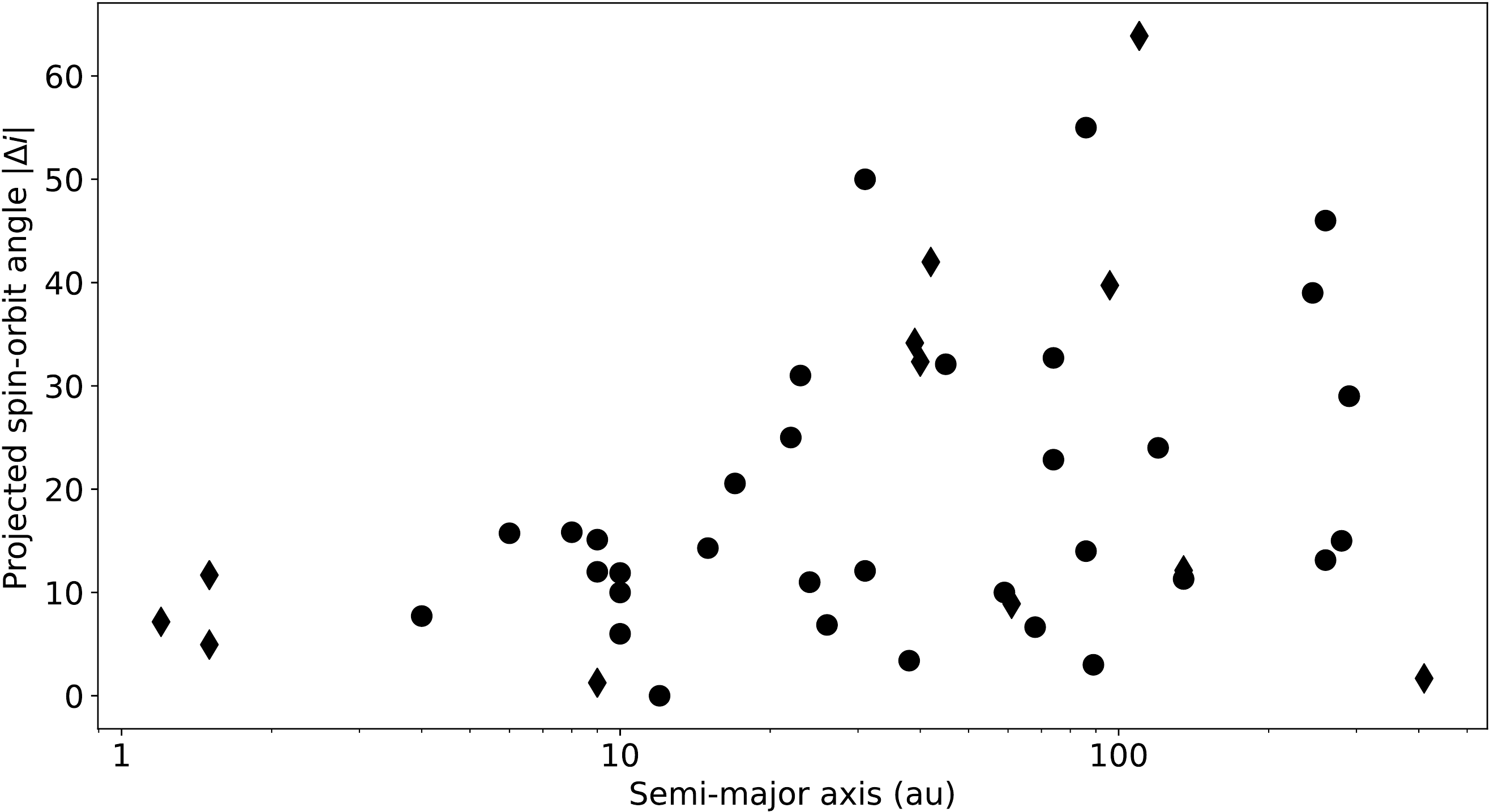}
      \caption{Reprint of Figure 2 from \citet{1994AJ....107..306H}, although excluding three wide systems without orbital solutions. Systems plotted with circles have known rotation periods from time series or calibrated activity indices. Systems plotted with diamonds have rotational velocities estimated from an average $v\sin i$ based on spectral class or by assuming a Skumanich relation $(v \propto \sqrt{\tau})$ calibrated with the Hyades and using age estimates from Ca II fluxes or simply assuming an age of 3\,Gyr. See \citet{1994AJ....107..306H} for details.}
         \label{fig:Hale_1994_Fig2_reprint}
\end{figure}

One of the most influential studies of spin-orbit alignment in binary stars was carried out by \citet{1994AJ....107..306H}, who used measurements of rotation periods and projected rotational velocities, $v\sin i$, to constrain stellar inclination angles in visual binaries with known orbital inclinations. \Hale\ found that binaries with semi-major axes smaller than $30-40\,$au are preferentially spin-orbit aligned while binaries in wider orbits are randomly aligned (see Fig.~\ref{fig:Hale_1994_Fig2_reprint}). \Hale\ found that while double star systems are aligned below a separation of $30-40\,$au, this trend is not true for hierarchical multiple systems where spin-orbit misalignment is seen at all separations. \Hale\ suggested that non-coplanarity may in fact decrease with separation in higher-order systems, the opposite trend seen for double star systems, although this trend is based on a small sample size.

\citet{2004RMxAC..21...15F} studied the spin-spin alignment of a large sample of $\sim 1000$ early-type binaries. They chose a sample of binaries (excluding triples or higher-order systems) in which they had measurements of $v\sin i$ of both components. They furthermore chose only main-sequence binaries of spectral type A9 or earlier. This was done to avoid slowly rotating stars or complications with loss of angular momentum during post-main-sequence evolution. They analysed the spin-spin alignment of the sample by comparing the $v\sin i$ of the two components in the binary with an artificially generated sample. They found that the stellar spin-axes of binary components are preferentially aligned at all orbital separations. The result of \citet{2004RMxAC..21...15F} was challenged by \citet{2009MNRAS.392..448H}, who analysed the same data and found that while the $v\sin i$ of components in close binaries with $a \lesssim 1\,$au appear to be correlated (which they attribute to tidal synchronisation), the data are insufficient to constrain the alignment of binaries with separations wider than $1\,$au (i.e.\ the data are compatible with both aligned or misaligned configurations). \citet{2012ApJ...750...79K} studied a smaller sample of 11 VLM (Very Low Mass, spectral type M7 or later) binaries with $a\lesssim 10\,$au. By analysing the $v\sin i$ of the individual components in the binaries, they find no evidence for preferential spin-spin alignment. They conclude that either the spin-axes of the components are misaligned or low-mass stars with similar spectral types rotate at significantly different speeds.

If a binary star is eclipsing, it is possible to constrain the projected obliquity by measuring the Rossiter-McLaughlin (RM) effect during the spectroscopic eclipse. The projected obliquity has been measured using the RM effect in just ten eclipsing binary systems\footnote{See \citet{albrecht_banana_2011-2} for an overview of quantitative and qualitative RM analysis results up to 2011. The RM effect has additionally been measured for a few low-mass and unequal-mass binary systems as part of the EBLM project \citep{2013A&A...549A..18T, 2019A&A...626A.119G, 2020MNRAS.tmp.2179H}, transiting brown dwarfs \citep{2012ApJ...761..123S} and hundreds of exoplanetary systems \citep[see \texttt{TEPCAT};][\url{www.astro.keele.ac.uk/jkt/tepcat/}]{2011MNRAS.417.2166S}.}. Here we briefly summarise the results: Five binaries were analysed by \citet{albrecht_spin_2007, albrecht_misaligned_2009-1, albrecht_banana_2011-2, albrecht_banana_2013, 2014ApJ...785...83A} as part of the BANANA (\textit{Binaries Are Not Always Neatly Aligned}) Project. The BANANA Project found that the massive B-type binaries DI Herculis \citep{albrecht_misaligned_2009-1} and CV Velorum \citep{albrecht_banana_2013} have large projected obliquities of both components. The remaining binaries in the BANANA Project all have well-aligned configurations. \citet{2011ApJ...741L...1W} measured the projected obliquity of the primary component in the planet-hosting eclipsing binary Kepler-16 and found the system to be aligned, possibly due to tidal realignment. Finally, \citet{2018MNRAS.478.1942S} measured the projected obliquities of four eclipsing binaries and reported alignment for all systems except a marginal detection of a misaligned secondary component of AI Phe, although more data are needed to confirm that result. They note that tidal forces are predicted to have realigned all but one of their binaries. DI Herculis and CV Velorum represent the only conclusive evidence for spin-orbit misalignment in close eclipsing binaries from the RM effect.

\citet{2013ApJ...776L..35Z} analysed the photometric light curve of KOI-368, an eclipsing binary with an A-star primary and a M-dwarf companion. By analysing the asymmetric eclipse of the gravity-darkened host star, they find that the $P = 110\,$d M-dwarf companion is in a significantly misaligned orbit.

There is evidence for both alignment and misalignment of protostellar and protoplanetary discs in binary systems. \citet{2004ApJ...600..789J} measured the polarisation of near-infrared light in young binary systems and found evidence that suggested protoplanetary discs in binaries with separations $200-1000\,$au are typically aligned to within $20^{\circ}$, although the authors could not eliminate the possibility that their results are influenced by interstellar polarisation. More recently, it has become possible to directly image the discs of wide binaries. These observations reveal that disc-disc, disc-orbit and circumbinary disc misalignment are commonly found \citep{2014Natur.511..567J, 2014ApJ...796..120W, 2016ApJ...830L..16B, 2018MNRAS.480.4738A, 2019NatAs...3..230K, 2020AJ....159...12M}. Measurements of the orientation of outflow axes have similarly revealed that protostellar jets in young multiple protostar systems are commonly misaligned with respect to each other \citep{2016ApJ...820L...2L}. 

\citet{2017ApJ...844..103T} investigated the orbital alignment in hierarchical triple systems and found that the orbits of systems with outer separations less than $\sim 50\,$au are preferentially aligned, while outer orbits with separations larger than $1000\,$au are misaligned with inner orbits. They note that the inner limit of $\sim 50\,$au roughly matches the size of the circumstellar disc, suggesting that dissipative interactions with the disc have influenced the formation or evolution of close triples, acting to align these systems. 

The results of \citet{1994AJ....107..306H} have been used to inform studies of binary formation and evolution. In particular, the preferential alignment of binaries with $a \lesssim 30\,$au has been used as evidence that close binaries form via disc fragmentation while wider binaries form via turbulent core fragmentation \citep[see e.g. reviews by][]{2011ASPC..447...47K, 2015ASPC..496...37B}. Recently, the \Hale\ study has received considerable attention in the context of exoplanetary systems due to the possible role of inclined binary companions in the formation of misaligned hot Jupiters by disc tilting and high-eccentricity migration.

The conflicting results for the spin-orbit alignment of binaries, the importance of this result for binary and planetary formation, and the recent developments in the consistent application of the $v\sin i$ method motivated us to revisit the seminal \citet{1994AJ....107..306H} study. We therefore obtained high-resolution spectra of binary stars in the \Hale\ study for a reanalysis of stellar parameters and projected rotational velocities. We reanalyse the \Hale\ sample following a Bayesian approach and compute posterior probability distributions of stellar inclination angles. In Section \ref{sec:Hale_Sample} we present the \Hale\ study and the sample of binary stars that we reanalyse in this work. In Section \ref{sec:Hale_reanalysis} we derive stellar parameters of the binary stars and compare our results to independent values. In Section \ref{sec:Hale_spinorbit_reanalysis_results} we present the projected spin-orbit alignment distribution of the binary sample based on our newly obtained data. In Section \ref{sec:Discussion} we compare the observed spin-orbit distribution to simulated observations with various measurement uncertainties. We additionally discuss systematic errors and possible biases in the binary sample. Finally, we summarise our findings in Section \ref{sec:Hale_conclusions}.

\section{The visual binary sample} \label{sec:Hale_Sample}
The \citet{1994AJ....107..306H} study focuses on solar-type binary and multiple systems with primary spectral types F5V-K5V. The spectral range was chosen to ensure a sample of binaries for which the projected rotational velocities $v\sin i$ and the rotational modulation could be reliably measured. Binary stars with orbits of at least grade 3 (1 best - 5 worst) in the Fourth Catalog of Orbits of Visual Binary Stars \citep[VB83;][]{1983PUSNO..24g...1W} were chosen. Binaries in hierarchical multiple systems were also included. The study was restricted to nearby systems with accurate distances such that the true value of the semi-major axis could be obtained. The systems were further selected to encompass a wide range of orbital separations ($1 - 10,1000\,$au) and eccentricities. The \Hale\ sample includes 86 stars in 73 stellar systems. 20 systems are triples or higher multiplicities. 

In this work, we focus exclusively on binary star systems from the \Hale\ sample (excluding stars in systems with higher multiplicities). We exclude higher-order systems due to the relatively small sample size. We further reject systems without orbital solutions (three systems) or known rotation periods (25 systems). Systems without known rotation periods are excluded due to concerns over the accuracy of rotation periods inferred from gyrochronology \citep[see e.g.][]{2016Natur.529..181V}. This leaves us with a sample of 35 stars in 29 double star systems.

We briefly summarise the stellar and orbital parameters used by Hale. \Hale\ compiled measurements of orbital elements, projected rotational velocities $v\sin i$, rotation periods $P_{\rm rot}$ and stellar radii $R_{\star}$ from original work and various sources in the literature. \Hale\ computed new orbits of six systems using measurements from literature combined with new astrometric observations. Hale combined literature values of $v\sin i$ with 32 new measurements. Stellar rotation periods were determined using one of three methods: rotational modulation of photometric light curves, time series analysis of spectroscopic Ca II fluxes, or calibration of chromospheric activity indices. The chromospheric flux ratio $R_{\rm HK}'$ (defined as the ratio of chromospheric emission in the cores of the Ca II H and K lines to the total stellar bolometric emission) is correlated with the stellar rotation period \citep{1984ApJ...279..763N}. A similar relation is seen for emission in Mg II lines. It is therefore possible to estimate stellar rotation periods from Ca II or Mg II fluxes by use of empirical activity-rotation relations. In the \Hale\ binary sample, 17 stars have rotation periods calibrated from their Ca II flux ratios, 14 stars have rotation periods from time series and four stars have rotation periods calibrated from their Mg II flux ratio. Rotation periods and activity indices were compiled by Hale from the literature. Finally, stellar radii were obtained by Hale from calibrated colour-radius relations or from literature. Hale derived stellar inclination angles and corresponding uncertainties by error propagation of the relation
\begin{align}
i = \arcsin{\left(\frac{v\sin i}{(2\pi R_{\star})/P_{\rm rot}}\right)}. \label{eq:inclination}
\end{align}
Systems with values of $\sin i$ greater than unity had their inclination angles fixed to $90^{\circ}$. This is the case for 19 of the 35 stars in the Hale binary sample.

\section{Analysis} \label{sec:Hale_reanalysis}
We reexamine the Hale study by applying a more statistically robust Bayesian formalism as well as determining the accuracy of the stellar parameters used in Hale's study using high-resolution spectroscopy. To establish the accuracy of the stellar parameters, we observed \nsong\ stars in the \Hale\ sample using the $1\,$m  robotic Hertzsprung SONG Telescope \citep{2014RMxAC..45...83A, 2019PASP..131d5003F}. We obtained spectra using the highest resolution configuration of $R=112,000$ of the SONG telescope's echelle spectrograph \citep{2017ApJ...836..142G}. Here we describe the derivation of stellar parameters from the SONG spectra and compare our results with Hale and other independent estimates.

\subsection{Deriving stellar parameters}
We analyse SONG spectra with the open-source Python tool \texttt{SpecMatch-emp} \citep{2017ApJ...836...77Y}.  \texttt{SpecMatch-emp} derives stellar parameters of FGKM stars by comparison of the stellar spectrum with an empirical spectral library of well-characterised stars. \texttt{SpecMatch-emp} derives stellar parameters with a typical precision of $100\,$K in stellar effective temperatures, $15\%$ in radii and $0.09\,$dex in metallicities. We analyse SONG spectra in the wavelength region $5000-5500\,$Å. 

\texttt{SpecMatch-emp} does not derive calibrated projected rotational velocities $v\sin i$. To derive $v\sin i$, we compute synthetic spectra with effective temperatures, surface gravities and metallicities as determined by \texttt{SpecMatch-emp}. We synthesise spectra using the stellar synthesis framework iSpec \citep{2014A&A...569A.111B} with the radiative transfer code SPECTRUM \citep{1994AJ....107..742G} using ATLAS9 atmospheres \citep{1979ApJS...40....1K, 2003IAUS..210P.A20C, 2005MSAIS...8...14K}, the VALD atomic line list \citep{2011BaltA..20..503K} and adopting \citet{2007SSRv..130..105G} solar abundances. We adopt the microturbulent velocity from the empirical relation implemented in iSpec. We do not apply instrumental, macroturbulent or rotational broadening at this stage, creating a sharp-lined template. Using our sharp-lined templates, we compute broadening functions (BFs) of each star \citep[see][]{1992AJ....104.1968R, 1999ASPC..185...82R, 2002AJ....124.1746R}. We fit the BF with a stellar line profile model including instrumental, macroturbulent and rotational broadening. We adopt Gaussian instrumental broadening corresponding to a resolution of $R = 112,000$ and macroturbulent broadening as determined from the empirical relation by \citet{2014MNRAS.444.3592D}. A few stars are slightly outside the $5200-6400\,$K temperature range of the \citet{2014MNRAS.444.3592D} macroturbulence relation. In such cases we extrapolate the relation. We derive projected rotational velocities $v\sin i$ from best-fit line profiles. The formal uncertainty on $v\sin i$ as derived by fitting a line profile model to the BF is not an accurate estimate of the true uncertainty. The uncertainties are underestimated partly because the uncertainty in the other stellar parameters are neglected. Imperfect normalisation may also affect the BF and potentially the $v\sin i$. By computing the BF of various stars in the sample using different sharp-lined templates, we have found that our $v\sin i$ are generally precise to $1\,$km/s or better.

Although the systems in our sample are visual binaries, some systems have small sky-projected separations. We therefore visually inspected the BF of each spectrum to look for double peaks or asymmetries which would indicate that the spectrum is double-lined or blended. We found seven stars with spectra that indicated binarity. Since \texttt{SpecMatch-emp} is not designed to handle blended spectra, the stellar parameters derived from double-lined SONG spectra are not used in the analysis. 

\begin{figure}
   \centering
   \includegraphics[width=\columnwidth]{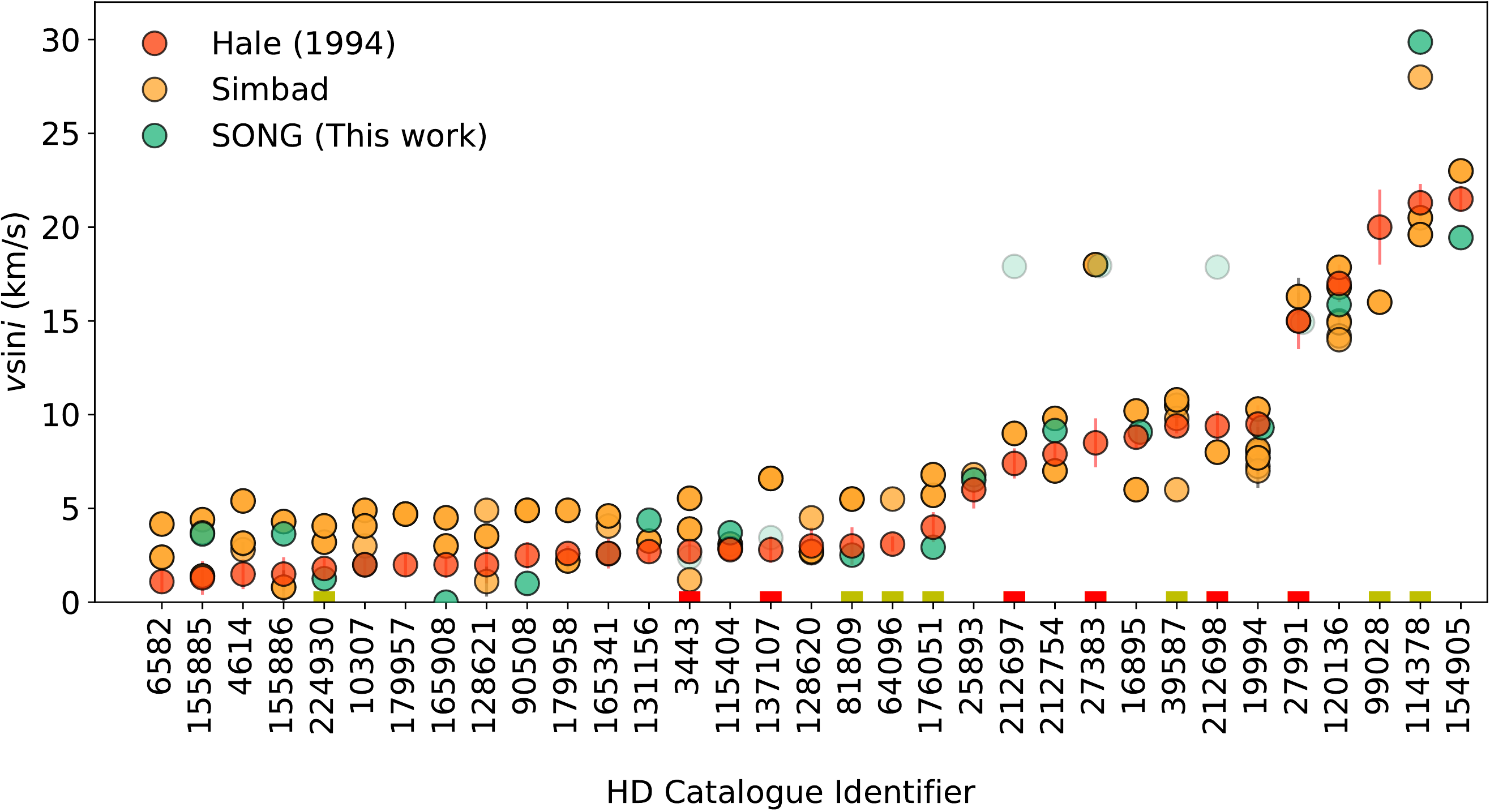}
      \caption{Comparison of projected rotational velocities $v\sin i$ from Hale, Simbad (a collection of published values), and SONG spectra (this work). Yellow squares indicate stars with bright, nearby companions. Red squares indicate stars identified in SONG spectra as double-lined. For visual clarity HD\,113139 at $v\sin i = 92\,$km/s is not shown. HD\,64096 does not have $v\sin i$ measurements from Simbad or SONG.}
         \label{fig:Hale_Simbad_SONG_vsinis}
\end{figure}

\begin{figure}
   \centering
   \includegraphics[width=\columnwidth]{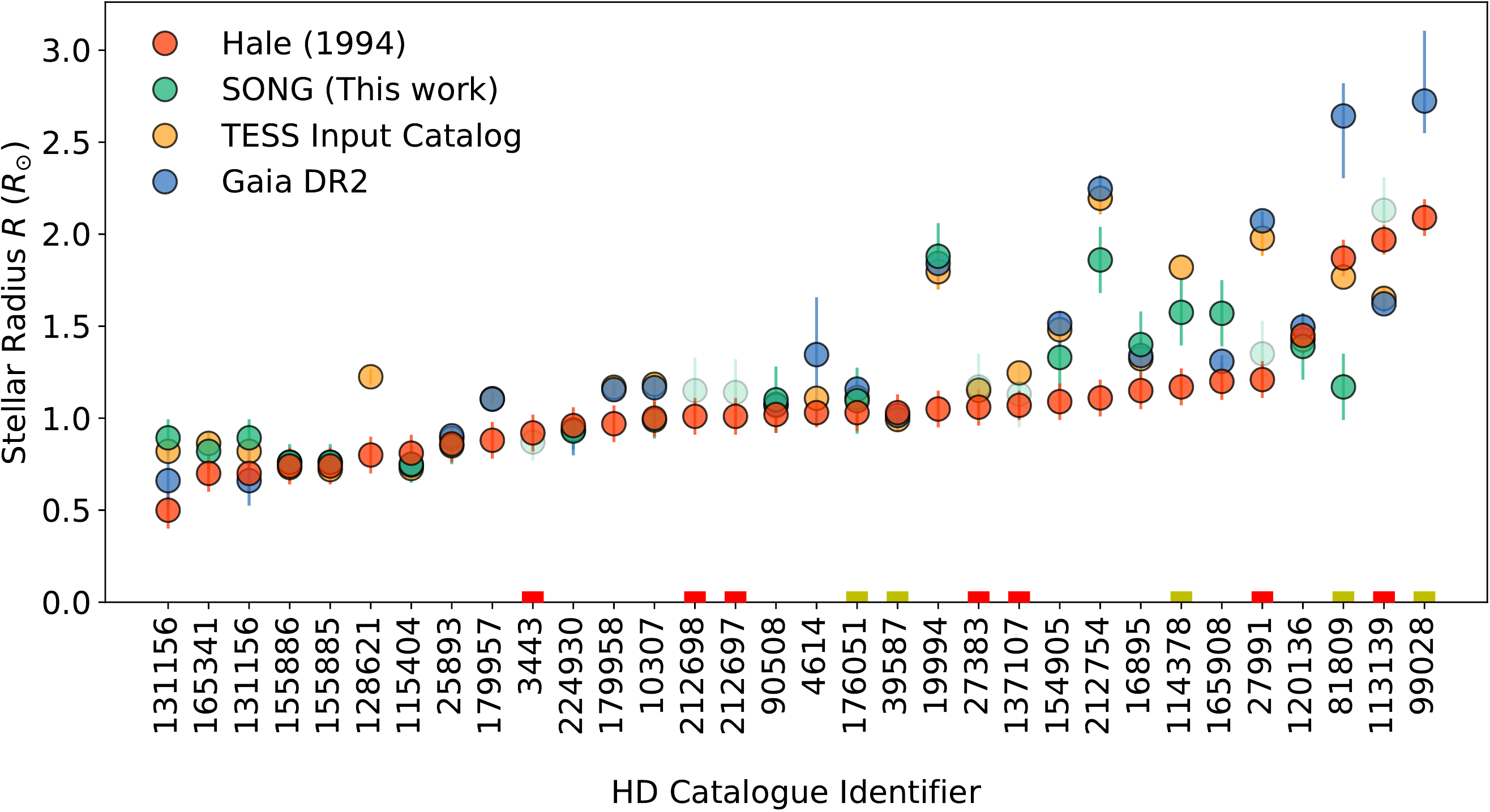}
      \caption{Comparison of stellar radii from Hale, $Gaia$ DR2, TESS Input Catalog, and SONG spectra (this work). Yellow squares indicate stars with bright, nearby companions. Red squares indicate stars identified in SONG spectra as double-lined. HD\,6582, HD\,64096 and HD\,128620 do not have radius estimates from SONG, \textit{Gaia} DR2 or TIC.}
         \label{fig:Hale_Gaia_TIC_SONG_Rs}
\end{figure}

\subsection{Comparison of projected rotational velocities}
In Fig.~\ref{fig:Hale_Simbad_SONG_vsinis} we plot a comparison between projected rotational velocities derived from SONG spectra, \Hale\, and various literature values. We note that we have included the derived parameters from double-lined spectra in Figs.~\ref{fig:Hale_Simbad_SONG_vsinis} and \ref{fig:Hale_Gaia_TIC_SONG_Rs} only to illustrate the effect of unmodelled binarity in the stellar spectra. The double-lined binaries HD\,3443 and HD\,137107 have clearly separated sets of spectral lines. For these stars we fitted both sets of lines in the BF and assumed that the deepest absorption lines belonged to the primary component. Since we study visual binaries, many stars will have nearby, bright companions that could affect the determination of stellar parameters. In Figs.~\ref{fig:Hale_Simbad_SONG_vsinis} and \ref{fig:Hale_Gaia_TIC_SONG_Rs}, we therefore mark stars with companions with projected separations less than $3.5"$ and magnitude differences less than $3.25$ ($>5\%$ flux difference) as listed in the Washington Double Star Catalog \citep[WDS;][]{2019yCat....102026M}. We mark systems visually identified as double-lined with red squares and stars with close companions in the WDS with yellow squares. We find overall good agreement with projected rotational velocities of the \nsingle\ single-lined binaries observed with SONG and Hale with a median difference of $0.01\pm 1.33\,$km/s (excluding the likely blended outlier HD\,114378). 
It is clear that at $v\sin i < 5\,$km/s, it is difficult to obtain fractional uncertainties better than $\sim 20\%$. At low rotational velocities, the rotational broadening, instrumental broadening, and macroturbulence all have similar magnitudes of a few kilometres a second. Literature values of $v\sin i$ are systematically larger than the $v\sin i$ derived in this work and Hale. This is likely due to neglect of macroturbulence in the literature values. The only significant disagreement between \Hale\ and SONG values is observed for HD\,114378, likely due to blended lines in the SONG spectrum (as discussed in Sec. \ref{ssec:Hale_notes_on_individual_systems}).

\subsection{Comparison of stellar radii}
In Fig.~\ref{fig:Hale_Gaia_TIC_SONG_Rs} we plot a comparison of stellar radii derived from SONG spectra, \Hale, \textit{Gaia} DR2 \citep{2016A&A...595A...1G, 2018A&A...616A...1G}, and the TESS Input Catalog \citep[TIC,][]{2019AJ....158..138S}. We crossmatched the \Hale\ sample with \textit{Gaia} DR2 by querying the \textit{Gaia} DR2 archive using HD Catalogue identifiers, which are resolved to coordinates and searched within a 3" radius. If multiple matches are found, we pick the brightest star within the search radius (since the stars in the \Hale\ sample are all nearby, bright stars). We checked that we selected the correct matches for binaries with both components within the search radius. We found 24 matches with stellar radii in the \textit{Gaia} DR2 archive. We similarly queried the TESS Input Catalog and found stellar radii for 26 stars in our sample. TESS and \textit{Gaia} DR2 radii are not independent since TESS radii are derived using \textit{Gaia} DR2 parallaxes with broadband photometry while \textit{Gaia} DR2 radii are derived using the parallax and \textit{Gaia} colours. We see good agreement for stars smaller than the Sun where SONG radii generally agree within $5\%$ or better with \textit{Gaia} DR2 and $10\%$ or better with \Hale. For a large fraction of stars larger than the Sun, there is significant disagreement between \Hale\ radii and independent estimates with \Hale\ radii being systematically smaller than other determinations. For stars larger than the Sun, SONG radii are generally accurate to within $\sim 15\%$ of the \textit{Gaia} DR2 radii, although some stars have larger disagreements. In Sec. \ref{ssec:Hale_notes_on_individual_systems}, we discuss these stars in detail.

\subsection{Adopted parameters and uncertainties}
We adopt $v\sin i$ and stellar radii from single-lined SONG spectra where available, supplemented by $v\sin i$ from \citet{1994AJ....107..306H} and radii from $Gaia$ DR2 or \citet{1994AJ....107..306H} for systems not observed by SONG or \textit{Gaia} DR2. For stars with significant disagreement between \Hale, SONG or \textit{Gaia} DR2, we review the literature to determine which value to adopt, see Sec. \ref{ssec:Hale_notes_on_individual_systems}.  Based on the comparison between independently derived values, we adopt uncertainties of $1.3\,$km/s on $v\sin i$, $5\%$ fractional errors on stellar radii smaller than the Sun and $15\%$ on radii larger than the Sun.

We adopt rotation periods from \citet{1994AJ....107..306H}. Nearly all stars in the sample have rotation periods estimated from time series of Ca II fluxes or via activity-rotation relations using the Ca II activity index $R_{\rm HK}'$. The Ca II fluxes used for these measurements were obtained at the Mount Wilson Observatory as part of the decades long survey started by \citet{1968ApJ...153..221W, 1978ApJ...226..379W} and continued by \citet{1980PASP...92..385V}. \citet{1984ApJ...279..763N} found that rotation periods estimated from $R'_{\rm HK}$ are generally accurate to $15\%$ or better. By analysing stars observed in multiple observing seasons, \citet{1996ApJ...466..384D} found that rotation periods derived from time series of Ca II fluxes are generally accurate to $\sim 15\%$. They attribute the scatter in rotation periods to the effect of surface differential rotation. We therefore adopt a fractional uncertainty of $15\%$ on all rotation periods. The data from the Mount Wilson survey are the longest-running and most comprehensive set of consistent observations available for the stars in the sample. It is therefore difficult to find independent estimates of the rotation periods of the stars in our sample. \citet{2010ApJ...725..875I} measured Ca II activity indices of 2630 stars at the Keck and Lick Observatories as part of the California Planet Search Program and estimated rotation periods using empirical activity-rotation relations. They observed six stars in our sample. We find a median difference of $25\%$ between the rotation periods of \Hale\ and \citet{2010ApJ...725..875I}, indicating that the uncertainty of rotation periods derived from  $R'_{\rm HK}$ could be larger than assumed, although this comparison is based on a very small sample size.

We use updated orbital inclinations from the Sixth Catalog of Orbits of Visual Binary Stars \citep[][ORB6]{2001AJ....122.3472H}. We find generally excellent agreement between orbital inclinations in ORB6 and \Hale\ (who used an earlier version of the same catalogue) with $\Delta i_{\rm orb} = 0.0^{+0.97}_{-1.3}$$^{\circ}$. Only five binaries differ by more than $5^{\circ}$ and no binary differ by more than $14^{\circ}$. We note that the good agreement between \Hale\ and ORB6 is not necessarily indicative that the orbital inclination is well-determined. The median orbital period of the sample is $152\,$yr. For many binaries, only few measurements have been added to the ORB6 catalogue since 1994. Due to the long orbital periods, the solutions are therefore expected to be similar. The ORB6 catalogue does not provide robust uncertainties on orbital elements. Following \Hale, we do not include the uncertainty of orbital inclinations in the analysis. The uncertainties on the projected spin-orbit angles therefore only reflect our knowledge of stellar inclination angles. We list adopted values, uncertainties, and references of all stars in the sample in Table \ref{tab:HALE_SAMPLE_adopted_parameters}.

\begin{sidewaystable*}
\centering
\caption{Adopted parameters of the binary sample.} \label{tab:HALE_SAMPLE_adopted_parameters}
\begin{tabular}{llllllllllll}
\hline\hline             
HD &
  $v\sin i$ &
  $\sigma(v \sin i)$ &
  $v\sin i$ Ref. &
  $R_{\star}$ &
  $\sigma(R_{\star})$ &
  $R_{\star}$ Ref. &
  $P_{\rm rot}$ &
  $\sigma(P_{\rm rot})$ &
  $P_{\rm rot}$ Ref. &
  $i_{\rm orb}$ &
  $i_{\rm orb}$ Ref. \\
 &
  \multicolumn{1}{l}{(km/s)}  &
  \multicolumn{1}{l}{(km/s)} &
   &
  \multicolumn{1}{l}{($R_{\odot}$)} &
  \multicolumn{1}{l}{($R_{\odot}$)} &
   &
  \multicolumn{1}{l}{(d)} &
  \multicolumn{1}{l}{(d)} &
   &
  \multicolumn{1}{l}{($^\circ$)} &
   \\
\hline
3443   & 2.7   & 1.3 & 1  & 0.92 & 0.05 & 1  & 32.6 & 4.89 & 1 & 65.9  & 2 \\
4614   & 1.5   & 1.3 & 1  & 1.35 & 0.2  & 3 & 16.5 & 2.48 & 1 & 35.6  & 2 \\
6582   & 1.1   & 1.3 & 1  & 0.78 & 0.12 & 1  & 29.8 & 4.47 & 1 & 73.2  & 2 \\
10307  & 1.99  & 1.3 & 4 & 0.99 & 0.05 & 4 & 22.1 & 3.32 & 1 & 81.2  & 2 \\
16895  & 9.08  & 1.3 & 4 & 1.4  & 0.21 & 4 & 6.7  & 1.0  & 1 & 75.44 & 2 \\
19994  & 9.31  & 1.3 & 4 & 1.9  & 0.07 & 5 & 19.0 & 2.85 & 1 & 65.9  & 2 \\
25893  & 6.52  & 1.3 & 4 & 0.85 & 0.04 & 4 & 7.51 & 1.13 & 1 & 76.0  & 2 \\
27383  & 8.5   & 1.3 & 1  & 1.06 & 0.16 & 1  & 5.4  & 0.81 & 1 & 53.0  & 2 \\
27991  & 15.0  & 1.3 & 1  & 1.16 & 0.06 & 6  & 2.9  & 0.44 & 1 & 58.0  & 2 \\
39587  & 9.4   & 1.3 & 1  & 1.01 & 0.15 & 3 & 5.25 & 0.79 & 1 & 84.06 & 2 \\
64096  & 3.1   & 1.3 & 1  & 1.13 & 0.17 & 1  & 22.3 & 3.34 & 1 & 80.82 & 2 \\
81809  & 2.51  & 1.3 & 4 & 2.64 & 0.4  & 3 & 40.5 & 6.08 & 1 & 85.4  & 2 \\
90508  & 1.0   & 1.3 & 4 & 1.1  & 0.16 & 4 & 18.0 & 2.7  & 1 & 81.4  & 2 \\
99028  & 20.0  & 1.3 & 1  & 2.928 & 0.103 & 7  & 1.8  & 0.27 & 1 & 52.0  & 2 \\
113139 & 92.0  & 1.3 & 1  & 1.62 & 0.24 & 3 & 0.8  & 0.12 & 1 & 46.9  & 2 \\
114378 & 21.3  & 1.3 & 1  & 1.58 & 0.24 & 4 & 3.01 & 0.45 & 1 & 89.95 & 2 \\
115404 & 3.7   & 1.3 & 4 & 0.75 & 0.04 & 4 & 18.9 & 2.83 & 1 & 86.59 & 2 \\
120136 & 15.87 & 1.3 & 4 & 1.39 & 0.21 & 4 & 5.2  & 0.78 & 1 & 37.0  & 2 \\
128620 & 3.0   & 1.3 & 1  & 1.23 & 0.18 & 1  & 25.1 & 3.76 & 1 & 79.32 & 2 \\
128621 & 2.0   & 1.3 & 1  & 0.8  & 0.12 & 1  & 25.1 & 3.76 & 1 & 79.32 & 2 \\
131156 & 4.37  & 1.3 & 4 & 0.89 & 0.04 & 4 & 6.14 & 0.92 & 1 & 41.0  & 2 \\
137107 & 2.8   & 1.3 & 1  & 1.07 & 0.16 & 1  & 13.6 & 2.04 & 1 & 58.08 & 2 \\
154905 & 19.45 & 1.3 & 4 & 1.33 & 0.2  & 4 & 2.75 & 0.41 & 1 & 37.8  & 2 \\
155885 & 3.64  & 1.3 & 4 & 0.76 & 0.04 & 4 & 21.5 & 3.22 & 1 & 80.21 & 2 \\
165341 & 2.58  & 1.3 & 4 & 0.82 & 0.04 & 4 & 19.5 & 2.92 & 1 & 58.9  & 2 \\
165908 & 0.0   & 1.3 & 4 & 1.57 & 0.24 & 4 & 9.1  & 1.36 & 1 & 36.18 & 2 \\
176051 & 2.93  & 1.3 & 4 & 1.1  & 0.16 & 4 & 17.5 & 2.62 & 1 & 64.2  & 2 \\
179957 & 2.0   & 1.3 & 1  & 1.1  & 0.17 & 3 & 25.2 & 3.78 & 1 & 60.94 & 2 \\
179958 & 2.6   & 1.3 & 1  & 1.16 & 0.17 & 3 & 26.9 & 4.03 & 1 & 60.94 & 2 \\
212697 & 7.4   & 1.3 & 1  & 1.01 & 0.15 & 1  & 5.8  & 0.87 & 1 & 44.13 & 2 \\
212698 & 9.4   & 1.3 & 1  & 1.01 & 0.15 & 1  & 7.8  & 1.17 & 1 & 44.13 & 2 \\
212754 & 9.15  & 1.3 & 4 & 1.86 & 0.28 & 4 & 11.1 & 1.66 & 1 & 87.0  & 2 \\
224930 & 1.25  & 1.3 & 4 & 0.93 & 0.05 & 4 & 23.8 & 3.57 & 1 & 49.0  & 2 \\
\hline
\end{tabular}
\tablebib{(1)~\citet{1994AJ....107..306H} and references therein; (2)~ORB6 \citep{2001AJ....122.3472H}; (3)~\textit{Gaia} DR2 \citep{2016A&A...595A...1G, 2018A&A...616A...1G};
(4)~This work (spectroscopic); (5)~\citet{van_Belle_2009}; (6)~ This work (isochrone); (7)~\citet{2006MmSAI..77..411T}}
\end{sidewaystable*}

\subsection{Notes on individual stars} \label{ssec:Hale_notes_on_individual_systems}
Here we discuss stars for which there is significant disagreement in the derived radii or projected rotational velocities from different sources. \\

\noindent \textbf{HD\,19994}\\
\noindent The Hale stellar radius at $R = 1.05\pm0.1\,R_{\odot}$ is significantly smaller than the radii of $R \sim 1.9\,R_{\odot}$ predicted by \textit{Gaia} DR2, TIC and this work. HD\,19994 has been analysed interferometrically and found to have a radius of $R = 1.898\pm0.070\,R_{\odot}$ \citep{van_Belle_2009}, in agreement with the analysis of the SONG spectrum. We adopt the interferometric radius.\\

\noindent \textbf{HD\,81809}\\
\noindent The spectroscopic SONG radius at $R = 1.17\pm0.18\,R_{\odot}$ is significantly smaller than other determinations at $R=1.8-2.6\,R_{\odot}$. HD\,81809 has a close companion within $0.1"$ at a magnitude difference of $\Delta m = 1.9$, potentially affecting values derived from both photometry and spectroscopy. The SONG spectrum does not appear double-lined from a visual inspection of the BF. The star is however listed as a spectroscopic binary on Simbad. Using resolved broadband photometry and the \textit{Gaia} DR2 parallax, \citet{Egeland_2018} finds that HD\,81809 is a sub-giant with a radius of $2.42\pm0.08\,R_{\odot}$, consistent with the \textit{Gaia} DR2 radius. We adopt the \textit{Gaia} DR2 radius. \\

\noindent \textbf{HD\,212754}\\
\noindent The \Hale\ radius of $R = 1.11\pm 0.1\,R_{\odot}$ is significantly smaller than other determinations of $1.9-2.2\,R_{\odot}$ . HD\,212754 was recently discovered to be a single-lined spectroscopic binary with a period of $2.5\,$ years \citep{2003ApJS..147..103G, Willmarth_2016}. Since the spectral lines of the companion are not visible it is unlikely to have affected the spectroscopic parameters. Based on the absolute visual magnitude, \citet{Willmarth_2016} conclude that the star is evolved, in agreement with the spectroscopic SONG radius of $R = 1.86\pm 0.28\,R_{\odot}$ which we adopt.\\

\noindent \textbf{HD\,27991}\\
\noindent The radius listed in \textit{Gaia} DR2 and TIC at $R\sim 2.1\,R_{\odot}$ is significantly larger than the \Hale\ and SONG radius of $R\sim 1.3\,R_{\odot}$. This star has a companion at $0.1"$ with a magnitude difference of $\Delta m = 0.74$.  The BF of the SONG spectrum is visibly asymmetric, indicating contamination from the companion. We therefore cannot trust the spectroscopic radius. \citet{2019A&A...630A..96A} analysed the binary system in a joint spectroscopic and astrometric analysis and derived a dynamical mass of $M = 1.218 \pm 0.04\,M_{\odot}$. They further find that the dynamical parallax is incompatible with the \textit{Gaia} DR2 parallax, leading to incorrect radius estimates in \textit{Gaia} DR2 and TIC. Unfortunately, the authors did not estimate the stellar radius. To determine the radius of HD\,27991, we therefore derive stellar parameters from an isochrone analysis. HD\,27991 is a member of the Hyades. We adopt an age of $625\pm50\,$Myr and metallicity of $0.14\pm 0.05\,$dex \citep{1998A&A...331...81P} and fit the dynamical mass, age and metallicity of HD\,27991 to a grid of BaSTI isochrones \citep{Hidalgo2018} using the Bayesian Stellar Algorithm \texttt{BASTA} \citep{2015MNRAS.452.2127S}. We derive a stellar radius of $R = 1.16\pm 0.06\,R_{\odot}$ in good agreement with the SONG and \Hale\ radii. The SONG metallicity [Fe/H]$=0.13\pm 0.09$ and temperature $T_{\rm eff} = 6344\pm 110\,$K are likewise in good agreement with the Hyades metallicity and isochrone-derived temperature of $6239\pm127\,$K, indicating that the derived spectroscopic parameters are not significantly affected by the companion. We adopt the isochrone-derived radius. \\

\noindent \textbf{HD\,114378}\\
\noindent The \Hale\ radius of $1.17\pm 0.1\,R_{\odot}$ is significantly smaller than the SONG and TIC radii of $R\sim 1.6-1.8\,R_{\odot}$. The SONG $v\sin i$ of $30\,$km/s is significantly larger than the \Hale\ value of $21\,$km/s. HD\,114378 has a companion at $\sim 0-0.5"$ (depending on the position in the $26\,$ year orbit) with $\Delta m = 0.68$. By close inspection, the BF of the SONG spectrum appear slightly asymmetric, indicating possible contamination from the companion. This slight asymmetry was missed in the initial visual inspection of the BF. \citet{2011ApJ...743...48W} derived a radius of $1.59\,R_{\odot}$ by isochrone-fitting of broadband photometry but did not report the uncertainty. This radius is in agreement with the SONG radius of $1.58\pm 0.18\,R_{\odot}.$  We therefore adopt the SONG radius and \Hale\ $v\sin i$.\\

\noindent \textbf{HD\,113139}\\
\noindent The \Hale\ radius of $\sim 2\,R_{\odot}$ is larger than the \textit{Gaia} DR2-based radius of $\sim1.6R_{\odot}$. HD\,113139 has a companion at $0.9"$ with $\Delta m = 2.86$. The BF of the SONG spectrum is clearly double-peaked. \citet{2019A&A...623A..72K} derived a radius of $2.84\pm 0.14\,R_{\odot}$ using a brightness-colour relation. This radius is in significant disagreement with other determinations and is likely affected by contamination from the companion. The \Hale\ radius is similarly derived from a colour-radius relation and may be contaminated. For this star we adopt the \textit{Gaia} DR2 radius of $R = 1.62\pm 0.24\,R_{\odot}$.\\

\noindent \textbf{HD\,99028}\\
\noindent The \Hale\ radius of $2.1\pm0.1\,R_{\odot}$ is significantly smaller than the \textit{Gaia} DR2 radius of $2.72^{+0.38}_{-0.17}\,R_{\odot}$. \citet{2006MmSAI..77..411T} measured an interferometric radius of $2.928\pm0.103\,R_{\odot}$, consistent with the \textit{Gaia} DR2 radius. We adopt the interferometric radius.\\

\noindent \textbf{HD\,155885 and HD\,155886}\\
\noindent HD\,155885 and HD\,155886 make up the bright binary 36 Oph. The two components are separated by $\sim 4"$ and are equally bright to a precision better than $1\%$ \citep[$V_T = 5.12\pm 0.01$,][]{2002A&A...384..180F}.  The two components have equal radii, $v\sin i$ and rotation periods within the uncertainties of \Hale\ and \textit{Gaia} DR2, strongly indicating that the binary is equal-mass. We attempted to observed both components of 36 Oph with SONG. However, the similar brightness and small separation of the two components confused the automatic target acquisition of the robotic telescope. A visual inspection of the two SONG spectra reveal that they are identical at the level of the S/N of the spectra. We therefore possibly observed the same target twice. It is however unclear if we observed HD\,155885 or HD\, 155886. Using the SONG spectrum, we derive a radius of $R = 0.76\pm 0.1\,R_{\odot}$ and $v\sin = 3.6\pm 1\,$km/s, in agreement with the stellar parameters of both HD\,155885 and HD\,155886 in \Hale\ and \textit{Gaia} DR2. Since the derived stellar parameters agree within the uncertainties in all cases, we simply adopt the SONG radius and $v\sin i$ as the stellar parameters of HD\,155855 and reject HD\,155886 from further study. \\

\section{Results} \label{sec:Hale_spinorbit_reanalysis_results}
Having obtained independent stellar parameters for the majority of the \Hale\ binary sample and established their accuracy, we now estimate stellar inclination angles of stars in this sample. We compute stellar inclination angles using the Bayesian formalism of \citet{2020AJ....159...81M}. We use normal likelihood functions of $v\sin i$ and construct likelihood functions of the rotational velocities $v$ from the normal distributions of stellar radii and rotation periods. We assume uniform priors in $\cos i$ and $v$. We compute posterior probability distributions of stellar inclination angles $p(\cos i|v, v\sin i)$ using Eq.~(10) in \citet{2020AJ....159...81M}. We compute projected spin-orbit angles as $\Delta i = i - i_{\rm orb}$ with uncertainties derived from the $\cos i$ posterior probability distributions. We note that absolute value of $\Delta i$ is arbitrary in the sense that there is no difference between $\pm \Delta i$.
\begin{figure}
   \centering
   \includegraphics[width=\columnwidth]{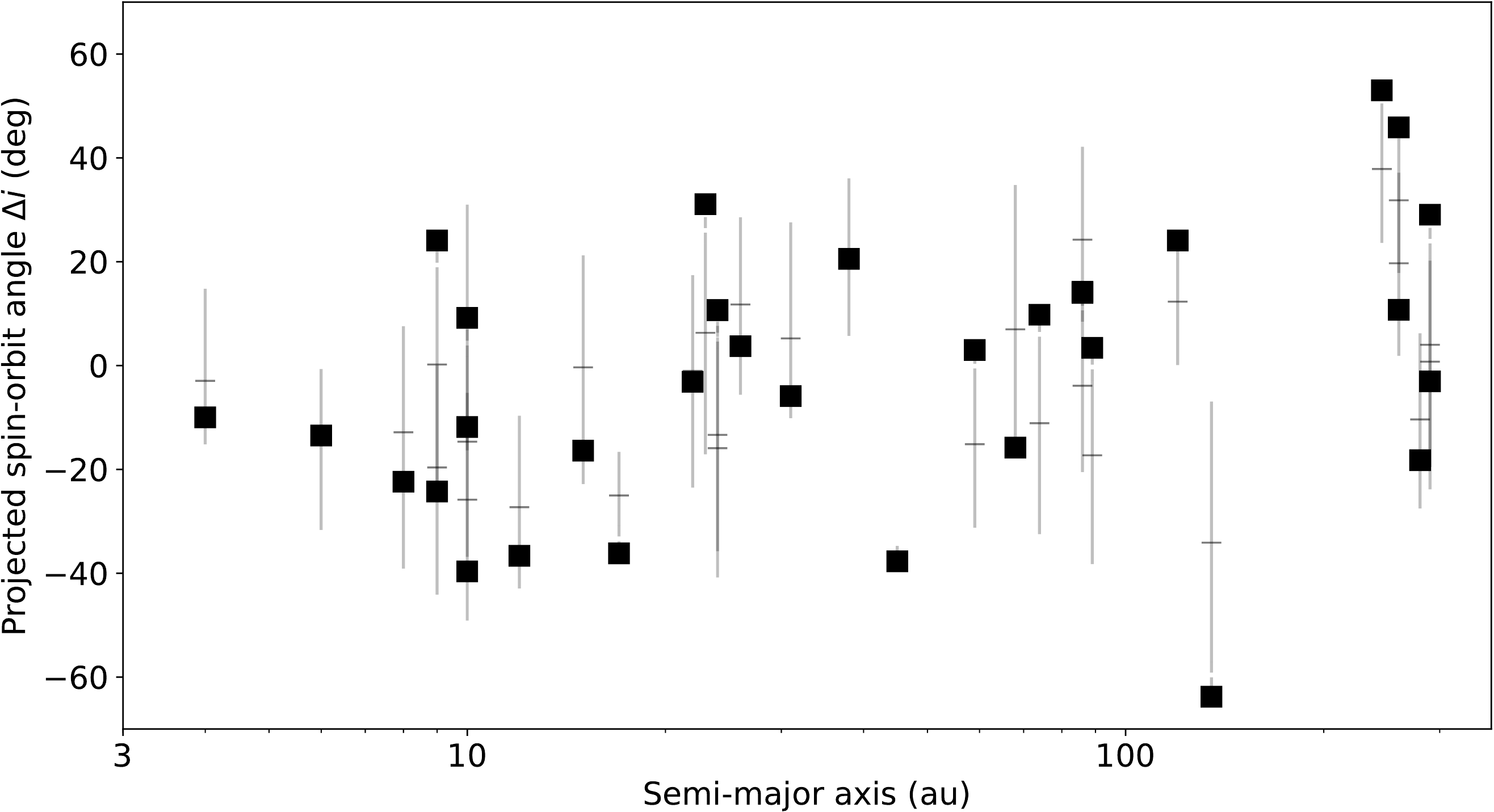}
      \caption{Projected spin-orbit alignment of solar-type binaries. Full vertical lines indicate $68\%$ confidence intervals. The median projected spin-orbit angle is marked with a short vertical line. The most likely projected spin-orbit angle (using the mode of the posterior) is indicated with a black square. For systems in which $\Delta i_{\rm mode}$ falls outside the $68\%$ confidence interval (typically if $i_{\rm mode} = 90^{\circ}$), a dashed line indicates the extension of the confidence interval.}
         \label{fig:Hale_deltai_new_params_new_analysis}
\end{figure}

In Fig. \ref{fig:Hale_deltai_new_params_new_analysis} we plot the projected spin-orbit angle $\Delta i$ as a function of semi-major axis. It is immediately apparent that we do not reproduce a clear difference between binaries with separations smaller and larger than $\sim 30\,$au. We find that nearly all stars in the sample have poorly constrained stellar inclination angles, with a median size of the $68\%$ confidence interval of $36^{\circ}$.

\section{Discussion} \label{sec:Discussion}
The difference between the $\Delta i$ distribution derived in this work (Fig. \ref{fig:Hale_deltai_new_params_new_analysis}) and Hale (Fig. \ref{fig:Hale_1994_Fig2_reprint}) is mainly a result of the different radii used in the two works. We found that some stars in the Hale study have smaller radii than the ones derived from high-resolution spectroscopy or \textit{Gaia} DR2 data. Another source of discrepancy comes from the derivation of stellar inclination angles. Hale derived stellar inclination angles by error propagation of Eq. \eqref{eq:inclination}, while we compute $\cos i$ posteriors using Bayesian statistics. As shown by \citet{2020AJ....159...81M}, simple error propagation (or simple Monte Carlo sampling) leads to incorrect $\cos i$ posteriors due to neglecting the correlation between stellar rotational velocity $v$ and projected rotational velocity $v\sin i$. The exclusion of stars with rotation periods estimated from gyrochronology is unlikely to have affected the conclusions. Of the 25 stars without known rotation periods in the Hale study, nine stars are members of higher-order systems and two stars were excluded by Hale for being unreliable. Below $30\,$au, only four stars without rotation periods were excluded from this analysis compared to the Hale study. These stars were weighted lower in the Hale analysis to account for their increased uncertainty. As demonstrated in the following section, a large sample of well-characterised stars are needed to make robust inferences about the spin-orbit distribution.

\subsection{Simulated observations} \label{ssec:Hale_simulations}
We compare the observed distribution of projected spin-orbit angles to the spin-orbit distribution of a randomly aligned population. This is done by generating samples of $\Delta i$ by drawing stellar and orbital inclination angles uniformly in $\cos i$. We compute the two-sided Kolmogorov–Smirnov (KS) statistic to test the hypothesis that the two distributions are drawn from the same underlying distribution\footnote{We note that the significance levels returned by the KS-test may not be reliable when comparing measurements with large uncertainties estimated from broad and asymmetric posteriors. We therefore caution against interpreting the $p$-values as precise estimators of significance.}. We find a $p$-value of $p = 0.4$, meaning that we cannot reject the null hypothesis with any significance.

We now test whether we could have detected if our binaries are all truly aligned. We simulate an aligned sample with characteristics similar to the \Hale\ sample using the following procedure: We assume that the observed stellar radii $R_{\star}$, rotation periods $P_{\rm rot}$, and orbital inclinations $i_{\rm orb}$ as listed in Table \ref{tab:HALE_SAMPLE_adopted_parameters} represent the true values of an aligned population of binaries. The $v\sin i$ of this aligned sample is computed as $v\sin i_{\rm orb} = (2\pi R_{\star})/P_{\rm rot}\sin i_{\rm orb}$. We then generate a set of observations by drawing values of $R_{\star}$, $P_{\rm rot}$, and $v\sin i_{\rm orb}$ using their estimated observational uncertainties as given in Table \ref{tab:HALE_SAMPLE_adopted_parameters} (using $\sigma_{v \sin i} = 1.3\,$km/s for $v\sin i_{\rm orb}$). Finally, using the artificially observed values (with the adopted observational uncertainties), we compute $\cos i$ posteriors and compute spin-orbit angles $\Delta i$ using the most likely stellar inclination angle $\Delta i = i_{\rm mode} - i_{\rm orb}$.

We simulate 100 well-aligned samples and compare each sample with a randomly aligned sample using the KS-test. We 
find that our artificially observed aligned samples are statistically indistinguishable to randomly aligned samples with a mean KS-statistic of $p = 0.3$. We therefore conclude that our data set is insufficient to make any conclusions about the spin-orbit alignment of binaries: Our data are compatible both with completely random alignment and with perfect alignment. With the adopted uncertainties of our $v\sin i$, radii and rotation periods, it is not possible to obtain stellar inclination angles of the required precision.

\begin{figure*}
   \centering
   \includegraphics[width=0.7\textwidth]{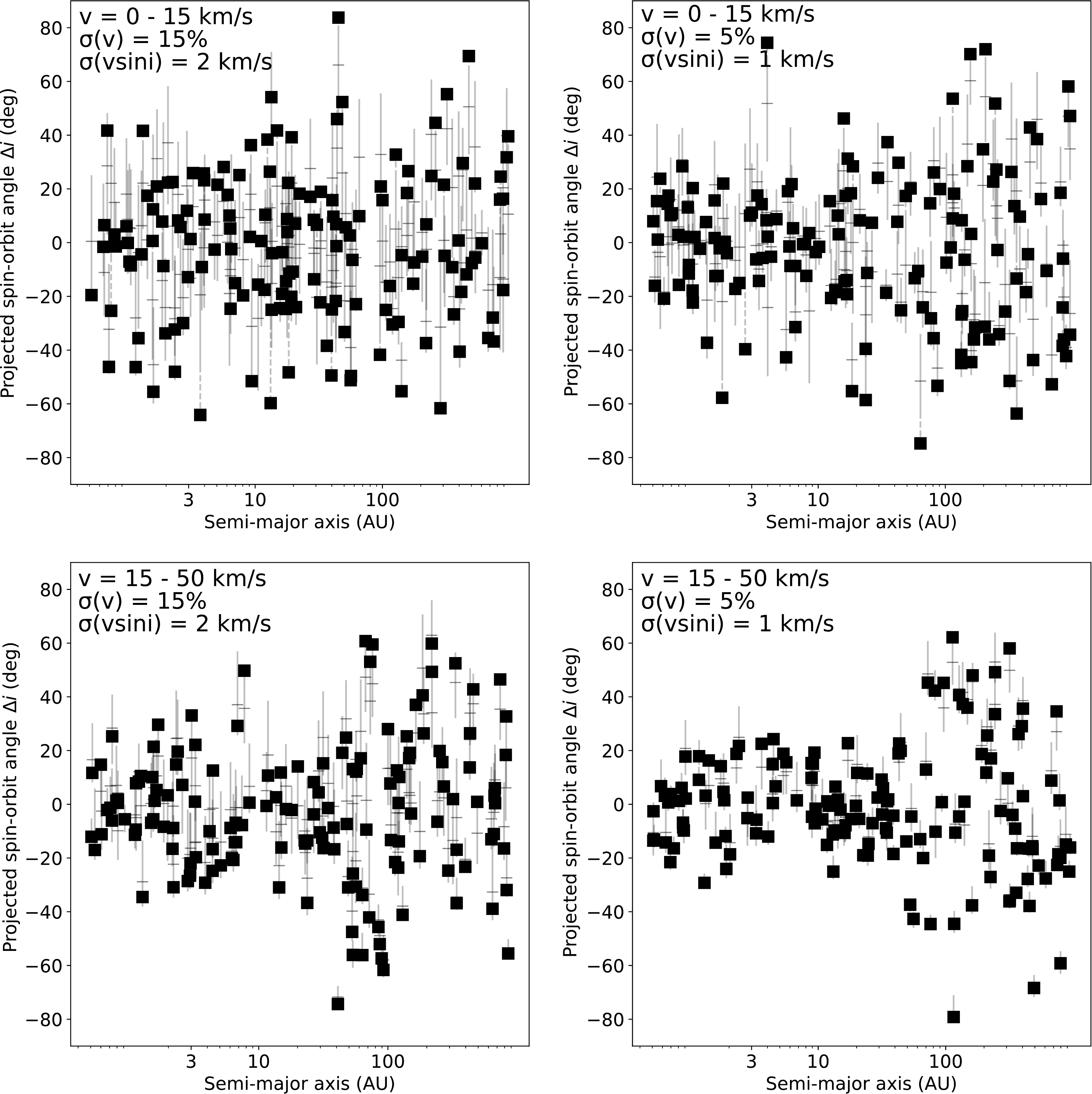}
      \caption{Projected spin-orbit distributions of simulated binary samples. Rotational velocities of binaries in the sample are drawn from a uniform distribution with ranges indicate in each panel. We describe $v$ and $v\sin i$ as normal distributions with measurement uncertainties indicated in each panel.}
         \label{fig:deltai_simulations}
\end{figure*}

Having found that the current sample cannot be used to learn about spin-orbit alignment in binary stars, what kind of sample is required? As a first step we now investigate the needed precision to distinguish between a  well-aligned and randomly aligned population. Even in a completely randomly aligned sample, the projected spin-orbit angle $|\Delta i|$ is less than $20^{\circ}$ for half of all systems and less than $60^{\circ}$ for 95$\%$ of systems. This is due to the fact that inclination angles are uniformly distributed in $\cos i$, not $i$. For stars with measurement uncertainties of $\sim 10\%$ on $v$ and $v\sin i$, the flat $\cos i$ posterior results in large uncertainties on stellar inclination angles, making it difficult to distinguish between preferentially aligned and misaligned populations without relatively large samples of precisely characterised systems. 

We simulated four binary samples, which we show in  Fig.~\ref{fig:deltai_simulations}. All simulated populations of binaries are preferentially aligned at small separations and randomly aligned at larger separations. We model the projected spin-orbit alignment distribution of binaries with separations less than $40\,$au as a normal distribution with a standard deviation of $10^{\circ}$. \footnote{We note that the choice of a normal distribution is somewhat arbitrary. \citet{2007ApJ...669.1298F} describes the spin-orbit distribution as a Fisher distribution. However, the exact choice of the distribution is not important here.} We simulate 150 binaries log-uniformly distributed over separations of $0.5-1000\,$au.  Orbital inclinations are assumed to be  uniformly distributed in $\cos i$. We do not include measurement uncertainty in orbital inclinations. We assign rotational velocities $v$ of binaries in our sample uniformly in the range $0-15\,$km/s or $15-50\,$km/s, roughly approximating a solar-like and a F-type sample. We adopt normal likelihood functions with fractional uncertainties of either $15\%$ and $5\%$ on rotational velocities and absolute uncertainties of $2\,$km/s or $1\,$km/s on $v\sin i$. The upper two panels of Fig.~\ref{fig:deltai_simulations} show the simulated solar-like sample. With uncertainties of $\sigma(v) = 15\%$ and $\sigma(v\sin i) = 2\,$km/s, we do not see any difference between close and wide binaries. Assuming optimistic uncertainties of $5\%$ on rotational velocities and $1\,$km/s on $v\sin i$, the projected spin-orbit distribution shows a hint of increased scatter at large separations (although simple KS- or Anderson–Darling tests do not establish statistically significant differences between the two populations). Increasing the rotational velocities of the sample (lower two panels of Fig. \ref{fig:deltai_simulations}) decreases the fractional uncertainty on $v\sin i$, thereby increasing the precision of the stellar inclination angles. With realistic uncertainties, the spin-orbit distribution similarly shows a hint of increased scatter at large separations, although with low significance. The final simulation (lower right panel) shows a fast rotating sample with optimistic uncertainties. In this case, the difference between close and wide binaries is finally clearly visible.

\subsection{Systematic uncertainties and biases} \label{sec:Hale_sys_unc_and_bias}
The derivation of the projected spin-orbit angle depends on precise and accurate measurements of the orbital inclination, stellar radius, stellar rotation period and projected rotational velocity. Each of these measurements may introduce difficult-to-quantify systematic uncertainties. The uncertainty on orbital inclination is neglected in this work. For wide binaries with sparsely sampled orbits, the uncertainty on the orbital inclination may be large.

Stellar rotation periods derived from chromospheric activity indices may be affected by a number of systematic uncertainties. Seasonal changes in the chromospheric activity of the star will affect the measured activity index. It is therefore important to use mean activity indices derived from measurements covering a large timeline. Other systematic error sources include large scatter in the empirical activity-rotation relation, stellar parameters outside calibrated range, differently calibrated activity indices, and stellar rotation affected by for example tidal spin-up or spin-down due to close companions. For rotation periods determined from time series analysis, there is a risk of identifying harmonics or aliases of the true rotational signal as the rotation period. For photometric rotation periods, non-eclipsing close binaries with ellipsoidal variations may be mistaken for rotational modulation. For hotter stars, there is a risk of confusing stellar rotation with stellar pulsations.

Projected rotational velocities depend on empirically calibrated macroturbulence. If the macroturbulence is incorrectly or inconsistently calibrated, $v\sin i$ values will be systematically under- or overestimated, at least for $v\sin i$ below a few km\,s$^{-1}$. Differential rotation has been neglected here but may affect the determination of both $v\sin i$ and $P_{\rm rot}$ to a small amount.

The derived stellar radius may be significantly off if the evolutionary state of the star is wrongly classified. The derivation of precise radii of binary components with small sky-projected separations is complicated by blended photometry, potentially incorrect parallaxes and blended absorption lines.

In addition to systematic uncertainties, the binary sample will be biased due to various selection criteria and observational constraints. Some systems may have undetected companions, either planetary or stellar. Such companions could have complicated the evolution of the system, potentially obscuring trends in the binary spin-orbit distribution. By comparing stellar and orbital inclination angles, it is only possible to constrain the projected alignment along the line of sight. Since the sky-projected alignment of the stellar spin is unconstrained, it is possible that some systems with small projected spin-orbit angles are misaligned in 3D space. It is similarly not possible to distinguish between prograde and retrograde orbits. 

A major difficulty is obtaining precise rotation periods for a range of spectral classes. Activity-rotation relations are only calibrated for solar-type stars. The correlation between rotation and chromospheric activity breaks down for stars hotter than $\sim 6200\,$K which have low chromospheric activity and shallow (or no) convective envelopes \citep{2008ApJ...687.1264M}. It is similarly more difficult to obtain photometric rotation periods of hot stars. Stars with shallow convective envelopes or radiative envelopes do not produce star spots similar to cooler stars with convective envelopes. It is however possible to measure rotation periods from other phenomena such as chemical spots (photometric or spectroscopic) or magnetic field inhomogeneities (spectroscopic or polarimetric), although these methods are not widely applicable. Solar-like stars with rotation periods determined from rotational flux modulation are biased towards inclinations near $90^{\circ}$ due to the inclination-dependent amplitude of the star spot signal. This bias does not apply to rapidly rotating and magnetically active stars which often have polar star spots \citep{schuessler_why_1992}.

\section{Conclusions}  \label{sec:Hale_conclusions}
We reanalysed a subset of the Hale sample using revised stellar and orbital parameters, partly based on new observations. We did not find evidence for the often-quoted statement that binary stars with separations less than $30\,$au are preferentially aligned or that binaries in wider orbits are preferentially misaligned. The trend observed in \citet{1994AJ....107..306H} is likely a result of underestimating the radii of some stars in the sample. Using a modern Bayesian formalism, we found that poorly constrained stellar inclination angles make inferences about the spin-orbit alignment of the current sample impossible (i.e. the data are compatible with both alignment and random orientation).

To date, only sparse observational data exist to constrain the spin-orbit alignment of binaries. Observations of a handful of protoplanetary discs in young binaries indicate that disc-disc misalignment may be common \citep{2014Natur.511..567J, 2016ApJ...830L..16B}. However, from measurements of polarised light there is similarly evidence for disc-disc alignment \citep{2004ApJ...600..789J}. We do not know how often tidal forces realign such protoplanetary discs or if the protoplanetary discs are misaligned with stellar spins \citep{2012Natur.491..418B}. From observations of the RM effect, two misaligned binaries and a dozen of aligned systems have been found \citep{albrecht_misaligned_2009-1, 2014ApJ...785...83A}. These observations cover only close binaries with periods less than $100\,$d. The orbital and tidal evolution of close systems are significantly different from the evolution of wider binaries, making it impossible to extrapolate from these results.

Based on simple simulated observations, we predict that it will require a large, homogeneous sample with precisely known orbital inclinations, orbital velocities and projected rotational velocities to investigate trends in the spin-orbit alignment distribution with the $v\sin i$ method. Such a sample does not currently exist and will be difficult to obtain in the near future. Photometric monitoring surveys \citep[e.g. TESS,][]{2015JATIS...1a4003R} and future \textit{Gaia} data releases may help in constraining rotation periods and orbital inclinations. By focusing on eclipsing binaries, the orbital inclination is constrained to $i_{\rm orb} \sim 90^{\circ}$, eliminating one source of uncertainty. This will however restrict the sample to systems with relatively close orbits. It is similarly possible to eliminate the orbital inclination by measuring the spin-spin alignment in visual binaries. An alternative route is the investigation of specific systems in greater detail, for example by identifying the best-studied systems or using alternative techniques such as direct imaging of discs \citep[e.g.][]{2014Natur.511..567J}, spectro-interferometry \citep[e.g.][]{LeBouquin+2009}, or measuring alignment via the RM effect \citep[e.g.][]{albrecht_spin_2007}. 

\begin{acknowledgements}
We thank Alan Hale for useful discussions and acknowledge his significant contributions in pioneering the $v\sin i$ technique by compiling and significantly expanding the sample of binary stars with estimated projected spin-orbit angles. Since 1994, the Hale study has remained the largest and most important study of the spin-orbit alignment of binary stars. We thank the referee for a quick and helpful response. Based on observations made with the Hertzsprung SONG telescope operated on the Spanish Observatorio del Teide on the island of Tenerife by the Aarhus and Copenhagen Universities and by the Instituto de Astrofísica de Canarias. We thank the SONG team members for their helpfulness and quick feedback on the manuscript. We acknowledge support from the Danish Council for Independent Research, through a DFF Sapere Aude Starting Grant no. 4181-00487B. Funding for the Stellar Astrophysics Centre is provided by The Danish National Research Foundation (Grant agreement no.: DNRF106). This research has made use of the Washington Double Star Catalog maintained at the U.S. Naval Observatory. This work has made use of data from the European Space Agency (ESA) mission
    {\textit Gaia} (\url{https://www.cosmos.esa.int/gaia}), processed by the {\textit Gaia}
    Data Processing and Analysis Consortium (DPAC,
    \url{https://www.cosmos.esa.int/web/gaia/dpac/consortium}). Funding for the DPAC
    has been provided by national institutions, in particular the institutions
    participating in the {\textit Gaia} Multilateral Agreement. This research made use of NumPy \citep{van2011numpy} and SciPy \citep{scipy}. This research made use of Astropy, a community-developed core Python package for Astronomy \citep{2018AJ....156..123T}. This research made use of matplotlib, a Python library for publication quality graphics \citep{matplotlib}. This research has made use of the SIMBAD database, operated at CDS, Strasbourg, France \citep{2000A&AS..143....9W}. This research has made use of NASA's Astrophysics Data System. 
\end{acknowledgements}

\bibliography{combined}
\bibliographystyle{aa}

\end{document}